\pdfoutput=1
\documentclass[3p]{elsarticle}

\usepackage[utf8]{inputenc}

\journal{Journal of the Mechanics and Physics of Solids}

\biboptions{authoryear}

\usepackage{color}
\usepackage{amsmath}
\usepackage{amsfonts}
\usepackage{bm} 
\usepackage{booktabs} 
\usepackage{array} 
\usepackage{paralist} 
\usepackage{verbatim} 
\usepackage{subfig} 
\usepackage[color=green!40]{todonotes}
\usepackage[colorlinks, citecolor=cyan, linkcolor=cyan]{hyperref}
\usepackage[utf8]{inputenc}

\newcommand{\rev}[1]{\textcolor[rgb]{0,0,1}{#1}}

\newcommand{\mb}[1]{\mathbf{#1}}

\begin{document}

\begin{frontmatter}
\title{Smart helical structures inspired by the pellicle of euglenids}
\author[aff1]{Giovanni Noselli\corref{cor1}}
\ead{giovanni.noselli@sissa.it}
\author[aff2,aff3]{Marino Arroyo}
\ead{marino.arroyo@upc.edu}
\author[aff1,aff4]{Antonio DeSimone}
\ead{desimone@sissa.it}
\address[aff1]{SISSA--International School for Advanced Studies, 34136 Trieste, Italy}
\address[aff2]{Universitat Polit\`ecnica de Catalunya--BarcelonaTech, 08034 Barcelona, Spain}
\address[aff3]{Institute for Bioengineering of Catalonia, The Barcelona Institute of Science and Technology, 08028 Barcelona, Spain}
\address[aff4]{The BioRobotics Institute, Scuola Superiore Sant'Anna, 56127 Pisa, Italy}
\cortext[cor1]{Corresponding author.}
\begin{abstract}
This paper deals with a concept for a reconfigurable structure bio-inspired by the cell wall architecture of euglenids, a family of unicellular protists, and based on the relative sliding of adjacent strips. Uniform sliding turns a cylinder resulting from the assembly of straight and parallel strips into a cylinder of smaller height and larger radius, in which the strips are deformed into a family of parallel helices. We examine the mechanics of this cylindrical assembly, in which the interlocking strips are allowed to slide freely at their junctions, and compute the external forces (axial force and axial torque at the two ends, or pressure on the lateral surface) necessary to drive and control the shape changes of the composite structure. Despite the simplicity of the structure, we find a remarkably complex mechanical behaviour that can be tuned by the spontaneous curvature or twist of the strips.
\end{abstract}
\begin{keyword}
\!\!\!euglenoid pellicle \sep helical bundles \sep morphing structures  \sep reconfigurable structures \sep  bio-inspired structures
\end{keyword}
\end{frontmatter}


\section{Introduction}\label{intro}

Structures capable of reconfiguring themselves, hence adapting their properties (shape, acoustic, mechanical, etc.) to changing demands on the functions they have to perform, are attracting increasing interest. Principles to achieve such reconfigurations include tunable deformations of gels \citep{klein_2007, kim_2012, lucantonio_2017a}, active polymers and nematic elastomers \citep{desimone_1999, warner_2003, van_oosten_2008, aharoni_2014, white_2015a,white_2015b,lucantonio_2017b}, electro-active materials \citep{shahinpoor_1998,bhattacharya_1999,ionov_2014}, fiber-reinforced natural or artificial composites \citep{erb_2012,wu_2013,bertoldi_2017}, pneumatic structures (e.g., McKibben pneumatic artificial muscles) or, more generally, pneumatic actuation in robotics \citep{Shepherd_2011, Tondu_2012, yang_2015}, buckling-induced rearrangement of the material micro-architecture \citep{bertoldi_2010}, origami and kirigami design principles \citep{filipov_2015,rafsanjani_2017,dias_2017,lipton_2018}, just to name a few.
One concrete example that has found practical applications  is deployable helical antennas for satellites, which can be folded to a compact structure to be housed in the payload bay of a launcher, and then unfolded to their working configuration as expanded helical antennas once they have reached their service location in space \citep{pellegrino_2001,pellegrino_2013}.
 
\begin{figure}[!th]
\centering
\includegraphics[width=0.95\textwidth]{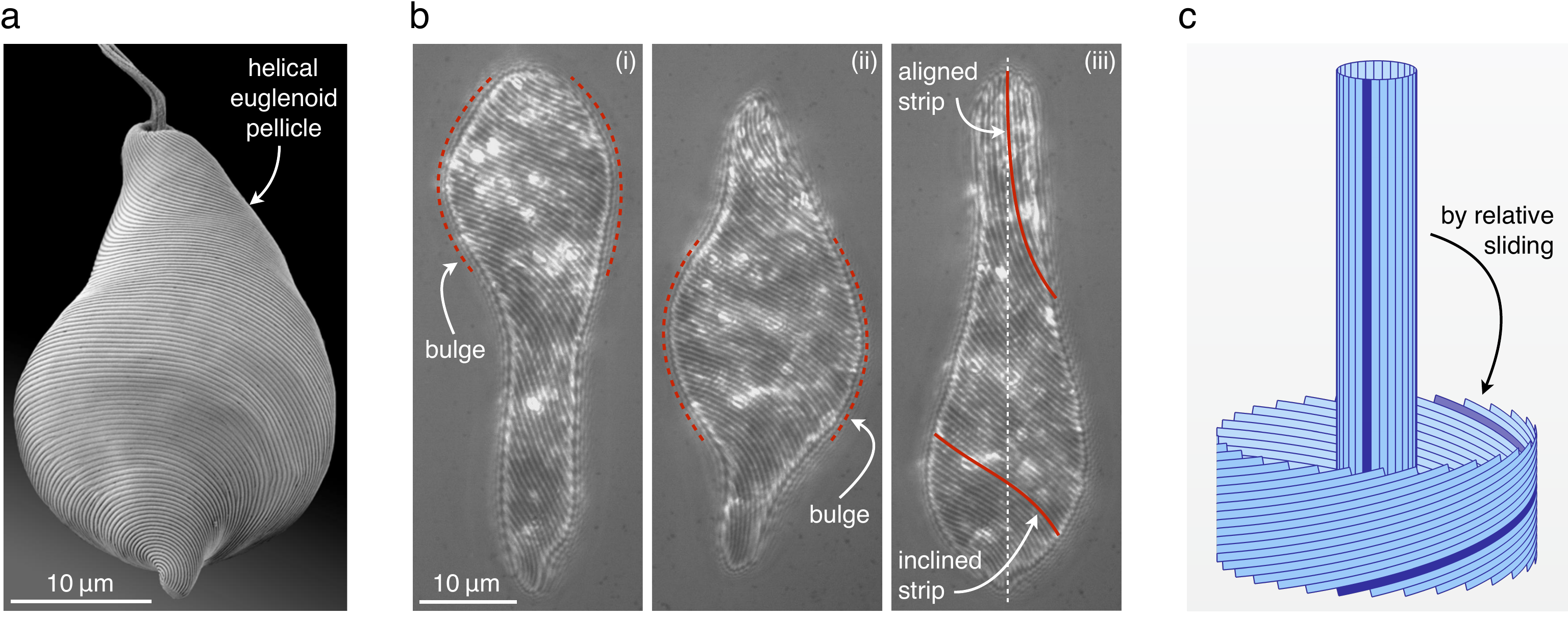}
\caption{(a) A scanning electron micrograph of {\it Eutreptia pertyi} showing helically arranged pellicle strips, adapted from \cite{leander_2001}. (b) Three micrographs of {\it Euglena gracilis} executing metaboly between a microscope slide and a cover slip. Observation by means of brightfield reflected light microscopy reveals the reconfiguration of the striated pellicle concomitant with cell body deformations. Micrographs were recorded at the SAMBA Lab of SISSA (c) Sketch of a structure with interlocking strips reminiscent of the shape morphing mechanism of the euglenoid pellicle. Adjacent strips are free to slide relative to each other at their junctions.}
\label{fig_1}
\end{figure}

Here, we draw inspiration in the pellicle of euglenids to discuss adaptive geometry and mechanical properties of an assembly of helical rods.  
Euglenids are a family of unicellular protists \citep{leander_2001,arroyo_2012, rossi_2017}, see Fig.~\ref{fig_1}a for a scanning electron micrograph of a sample of {\it Eutreptia petry} showing helically arranged pellicle strips.
Their unique shape-morphing principle based on the sliding of adjacent pellicle strips has been discussed in \cite{suzaki_1985, suzaki_1986}, and further explored in \cite{ arroyo_2012} and in \cite{arroyo_2014}.
This mechanism is most eloquently demonstrated by direct inspection of the behaviour of the biological template, see Fig.~\ref{fig_1}b, where in-vivo visualisation of 
sliding pellicle strips between microscope slides was achieved by exploiting bright-field reflected light microscopy.
%
%

Unlike other mechanisms, the one based on sliding of adjacent pellicle strips allows for very large local strains and shape changes. The resulting kinematics have been previously examined in a continuum limit, where we mapped the broad families of shapes --axisymmetric or not-- accessible by this mechanism, but not the underlying mechanics \citep{arroyo_2012,arroyo_2014}. In cells, shape regulation is provided by the activity of a large number of molecular motors  that exert forces on the microtubules present in the overlap region between two adjacent strips, causing them to slide.
In a biomimetic material inspired by the pellicle, it is unclear how the elasticity of the strips will control the overall mechanical properties of the assembly. This is the focus of the present paper, where we address the problem of computing the external equilibrium forces (axial force, axial torque, pressure on the lateral walls) necessary to drive and control the shape changes. This is studied for the restricted case of uniform cylindrical deformations, see Fig.~\ref{fig_1}c, and neglecting edge effects at the end of the structure and inter-strip friction, but fully retaining the geometric nonlinearities and the discreteness of the structure, containing a finite number of strips. Even in this restricted case, the system displays a highly nonlinear response, with interesting consequences for its mechanics.

The paper is organized as follows. We present in section~2 a detailed analysis for the kinematics and elasticity of cylindrical assemblies of interlocking rods capable of relative sliding at their junctions. We next examine in section~3 the mechanics of such assemblies for the loading cases of axial force along the cylinder axis, pure torque about that axis, and pressure acting on the lateral surface of the system. Finally, in section~4 we draw our conclusions and indicate possible future directions.

\section{Mechanics of helical assemblies of interlocking elastic rods}\label{sec_1}

\subsection{Kinematics}

\begin{figure}[!th]
\centering
\includegraphics[width=0.95\textwidth]{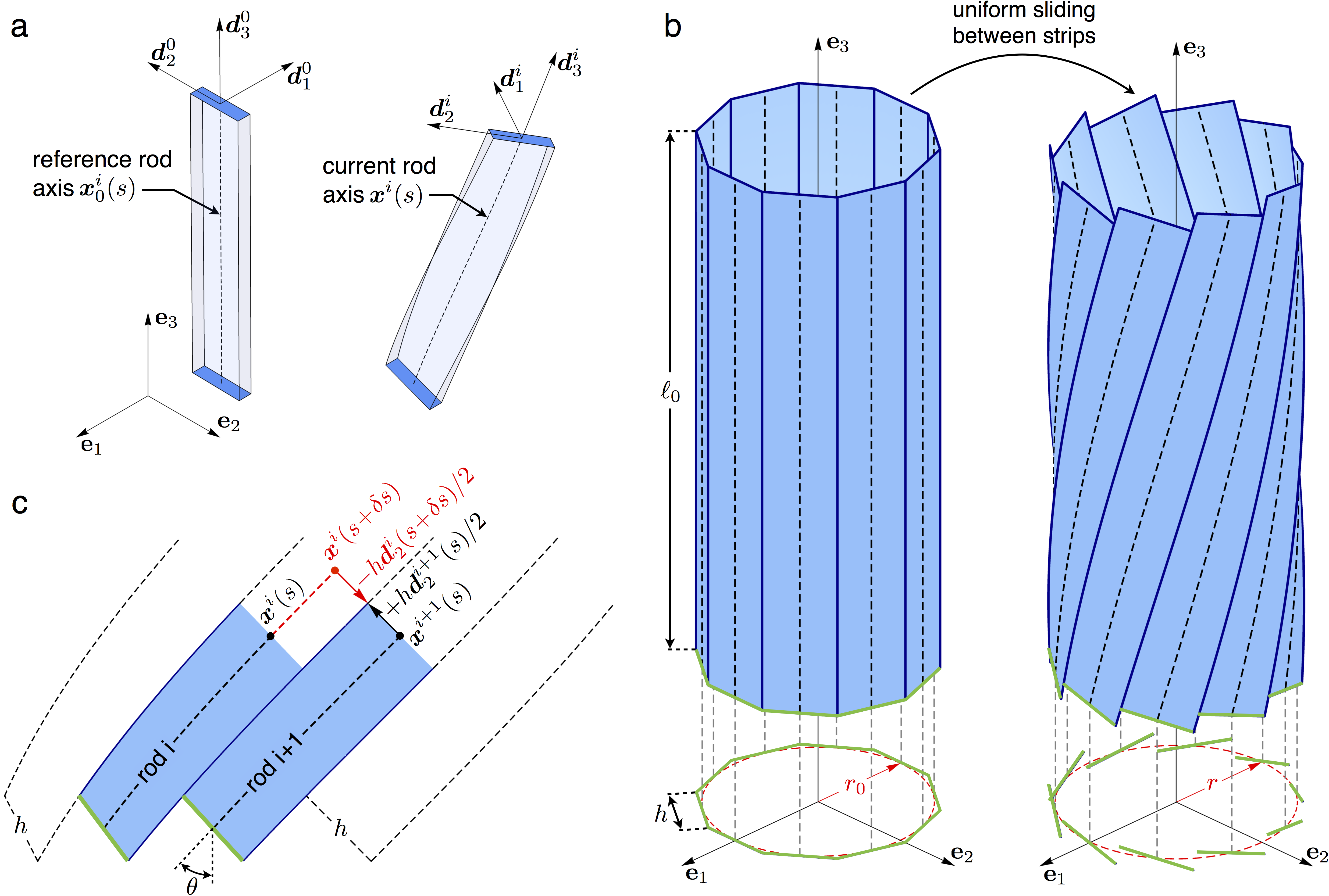}
\caption{(a) A inextensible and unshearable rod, straight in its reference configuration. The rod axis is denoted by $\bm{x}^i(s)$, whereas the triplet of directors $\{\bm{d}^i_1, \bm{d}^i_2, \bm{d}^i_3\}$ is introduced to characterize the current configuration of the rod. (b) A cylindrical assembly of rods in its reference and current configurations. Deformation of the system is achieved by the uniform relative sliding between adjacent strips. (c) A sketch of two adjacent strips undergoing relative sliding one to another. }
\label{fig_2}
\end{figure}

Inspired by the architecture of the euglenoid pellicle, we explore the response under mechanical loading of structural assemblies of interlocking strips capable of relative sliding at their junctions. We restrict our study to axisymmetric cylindrical geometries, in which the relative sliding between adjacent strips is constant along their edges \citep{arroyo_2014}, see Fig.~\ref{fig_1}c. We anticipate that,
under such assumption, a family of cylinders of increasing radius and decreasing height is obtained from a cylinder in which all the strips are initially aligned with the vertical direction, the same of the cylinder axis shown in Fig.~\ref{fig_2}b-left.

To account for the discrete nature of such assemblies, we model individual strips as inextensible and unshearable rods of constant cross section. Specifically, we denote by $\bm{x}^i(s)$ the parameterization of the $i$-th rod axis as a function of arc-length $s \in [0,\ell_0]$, see Fig.~\ref{fig_2}a. Here, $\ell_0$ is the rod length and corresponds to the height of the structural system in its reference configuration, Fig.~\ref{fig_2}b, that is, when the strips are straight and the relative sliding between them is null.
%
%
Adopting classical rod's theory \citep{antman_2005}, we characterize the current configuration of the rod by introducing an orthonormal triplet of directors $\{\bm{d}^i_1, \bm{d}^i_2, \bm{d}^i_3\}$, Fig~\ref{fig_2}a. In particular, $\bm{d}^i_1(s)$ and $\bm{d}^i_2(s)$ can be interpreted as characterizing the material cross section of the rod singled out by $s$, whereas $\bm{d}^i_3(s) = \bm{d}^i_1(s) \times \bm{d}^i_2(s)$ is the tangent to the rod axis, where \lq$\times$' denotes the vector product. \rev{Consequently, material cross sections will deform rigidly in three-dimensional space. We point out in passing that the modelling assumptions above are relevant to quasi-one-dimensional bodies in which all dimensions of the cross section are comparable. In the context of the present study, these conditions are met when the number of strips comprising the assembly is sufficiently large: as is intuitively clear, the width of the strips in the assembly decreases as the number of strips increases, see Eq.~\eqref{h_n}.  Different modelling approaches, based for instance on thin plate theory \citep{dias_2014}, could be more appropriate in other circumstances to account for the actual deformation of the cross section. 
}

The kinematics of the structural system follows from the assumption that adjacent strips will be compatible along their edges, both in the reference and in the current configuration, Fig.~\ref{fig_2}b. Denoting by $h$ the height of the cross section of each rod in the assembly, we parameterize the edges of the $i$-th rod as $\bm{x}^i_\pm(s) = \bm{x}^i(s) \pm (h/2) \bm{d}^i_2(s)$, where the subscripts \lq$+$' and \lq$-$' refer to the two edges of the rod, the left one and to the right one, respectively, in Fig.~\ref{fig_2}c.

Having restricted the study to axisymmetric cylindrical geometries, in the reference configuration the axis of each rod will lie on a cylindrical surface of radius $r_0$, see Fig.~\ref{fig_2}b-left. In view of the compatibility between adjacent edges, the first and the second director will be normal and tangent to that cylindrical surface, respectively. Consequently, the structural assembly will also be cylindrical and with a regular polygon circumscribed about a circle of radius $r_0$ as its cross section, see Fig.~\ref{fig_2}b-left. Denoting by $n$ the number of rods in the assembly, the radius $r_0$ and the height $h$ of the rods' cross section are related by
\begin{equation}
h = 2 r_0 \tan\left(\frac{\pi}{n}\right) .
\label{h_n}
\end{equation}

Upon deformation, initially straight strips will slide relative to adjacent ones, while their axes will bend and twist into circular helices wrapped around a cylindrical surface of radius $r$ and characterized by helix angle $\theta$, see Fig.~\ref{fig_2}b-right and Fig.~\ref{fig_2}c. Also in the current configuration, the first and the second director will be normal and tangent to that cylindrical surface, respectively. In general, $r$ will differ from $r_0$ and its actual value will depend upon the angle $\theta$. Our next goal is to determine the dependence of $r$ on $\theta$.

With $\{\mb{e}_1, \mb{e}_2, \mb{e}_3\}$ an orthonormal basis for the three-dimensional Euclidean space $\mathbb{E}^3$, the arc-length parameterization for the cartesian coordinates of the $i$-th rod axis $\bm{x}^{i}(s)$ reads
\begin{equation}
\bm{x}^{i}(s) = \{r \cos\left[2\pi (i-1)/n + (s/r) \sin \theta \right], r \sin\left[2\pi (i-1)/n + (s/r) \sin \theta \right], s \cos \theta\} , ~~~ s \in [0,\ell_0] ,
\label{rod_axis}
\end{equation}
where $s$ is the arc-length and $i\!=\!\{1, 2, \dots, n\}$ is an index denoting the rod number. Consequently, the tangent $\bm{t}^i(s)$ to the $i$-th rod axis reads
\begin{equation}
\bm{t}^i(s) = \{-\sin\theta \sin\left[2\pi (i-1)/n + (s/r) \sin \theta \right], \sin\theta \cos\left[2\pi (i-1)/n + (s/r) \sin \theta \right], \cos \theta\} , ~~~ s \in [0,\ell_0] ,
\label{tangent}
\end{equation}
whereas for the normal $\bm{n}^i(s)$ we compute
\begin{equation}
\bm{n}^i(s) = \{- \cos\left[2\pi (i-1)/n + (s/r) \sin \theta \right], - \sin\left[2\pi (i-1)/n + (s/r) \sin \theta \right], 0\} , ~~~ s \in [0,\ell_0] ,
\label{normal}
\end{equation}
such that the binormal $\bm{b}^i(s) = \bm{t}^i(s) \times \bm{n}^i(s)$  reads
\begin{equation}
\bm{b}^i(s) = \{\cos \theta \sin\left[2\pi (i-1)/n + (s/r) \sin \theta \right], -\cos \theta \cos\left[2\pi (i-1)/n + (s/r) \sin \theta \right], \sin \theta \} , ~~~ s \in [0,\ell_0] .
\label{binormal}
\end{equation}
We recall in passing that for a circular helix of radius $r$, curvature and torsion are given by $\kappa = \sin^2\theta/r$ and $\tau = \sin \theta \cos \theta / r$, respectively. 

For a given value of the helix angle $\theta$, the relation between the radii $r_0$ and $r$ can be determined by imposing a condition of kinematic compatibility between any pair of adjacent strips. We impose this condition at the junction of the strips, where relative sliding occurs. Recalling that $\bm{x}^i_\pm(s)$ denote the two opposite edges of the $i$-th rod, with reference to Fig.~\ref{fig_2}c we write
\begin{equation}
\bm{x}^i_{-}(s+\delta s) = \bm{x}^{i+1}_{+}(s) ,
\label{compatibility_1}
\end{equation}
from which we get
\begin{equation}
\bm{x}^i(s+\delta s) - (h/2) \bm{d}_2^i(s + \delta s) = \bm{x}^{i+1}(s) + (h/2) \bm{d}_2^{i+1}(s) ,
\label{compatibility_2}
\end{equation}
where $\delta s$ is a measure of the sliding between the axes of two consecutive strips, to be determined later. The compatibility condition of~\eqref{compatibility_2} provides a system of three scalar equations in the unknowns $\delta s$ and $r$. To proceed, we next notice that $\bm{d}^i_1(s) = \bm{n}^i(s)$, and that $\bm{d}^i_3(s) = \bm{t}^i(s)$, so that $\bm{d}^i_2(s) = \bm{b}^i(s)$. In view of Eqs.~\eqref{rod_axis}--\eqref{binormal}, one obtains that
\begin{equation}
\delta s = h \tan \theta ,
\label{deltas}
\end{equation}
and that
\begin{equation}
r = \frac{h}{2} \cot\left(\frac{\pi}{n} - \frac{h}{2r} \sin \theta \tan \theta \right) \cos \theta .
\label{r_theta}
\end{equation}
Substituting Eq.~\eqref{h_n} into Eq.~\eqref{r_theta}, we finally obtain the following relation for the dimensionless radius $\rho = r/r_0$ as a function of helix angle $\theta$ and number of rods $n$
\begin{equation}
\rho = \tan\left(\frac{\pi}{n}\right) \cot\left(\frac{\pi}{n} - \frac{1}{\rho} \tan\left(\frac{\pi}{n}\right) \sin \theta \tan\theta \right)\cos \theta .
\label{rho_theta}
\end{equation}

It is interesting to notice that Eq.~\eqref{rho_theta} simplifies into $\rho = 1/\cos\theta$ in the limit of $n \to \infty$. In such a limiting situation, the kinematics of the structural assembly can be described by its continuous approximation by exploiting the theory of non-Euclidean plates subject to a deformation field of pure shear along strips junctions \citep{efrati_2009, arroyo_2014}.
%
\begin{figure}[!th]
\centering
\includegraphics[width=0.525\textwidth]{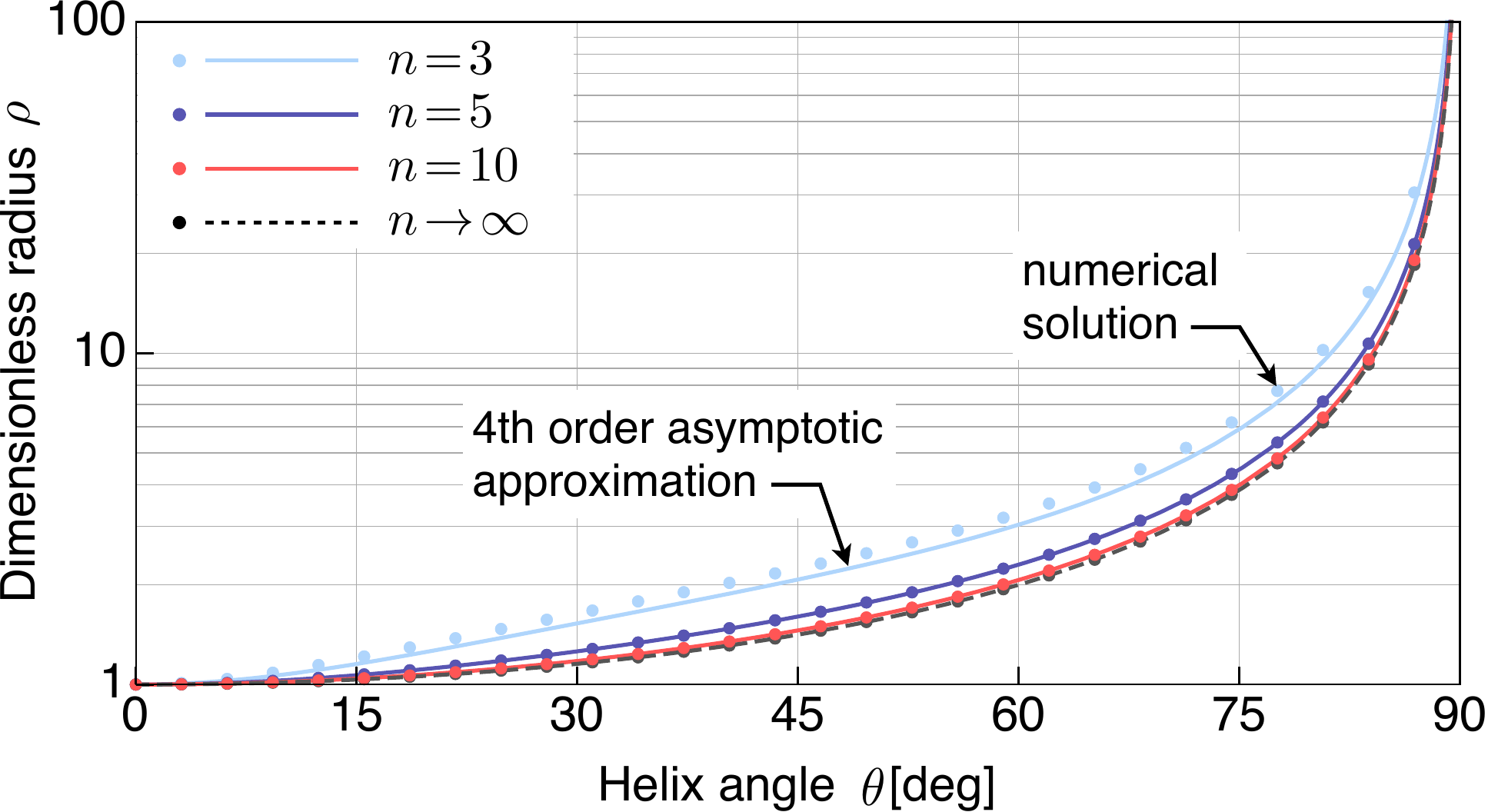}
\caption{Dimensionless radius $\rho = r/r_0$ as a function of helix angle $\theta$ for representative values of $n$ = \{3, 5, 10, $\infty$\}. Results (dots) from the numerical solution of equation~\eqref{rho_theta} are compared with the fourth-order asymptotic approximation of $\rho$, showing good agreement.}
\label{fig_3}
\end{figure}
%

The equation above provides the current, dimensionless radius as a function of helix angle in implicit form only. Nevertheless, Eq.~\eqref{rho_theta} can be solved numerically for $\rho$ at given values of $n$ and for $\theta \in [0, \pi]$. Also, an approximate solution can be readily obtained by means of an asymptotic expansion of $\rho$ in the smallness parameter $\epsilon = \pi/n$.\footnote{
In particular, by writing $\rho = \rho_0(\theta) + \epsilon \rho_1(\theta) + \epsilon^2 \rho_2(\theta) + \epsilon^3 \rho_3(\theta) + \epsilon^4 \rho_4(\theta) + \mathcal{O}(\epsilon^5)$, and by Taylor expanding the RHS of Eq.~\eqref{rho_theta} in the smallness parameter $\epsilon$ up to order four, one obtains
\begin{equation*}
\left\{
\begin{aligned}
	\rho_0(\theta) &= 1/ \cos\theta , \\
	\rho_1(\theta) &= 0 , \\
	\rho_2(\theta) &=  \sin \theta \tan \theta [15 + 8 \cos(2 \theta) + \cos(4 \theta)]/24 , \\
	\rho_3(\theta) &= 0 , \\
	\rho_4(\theta) &= \sin \theta \tan \theta [3660 + 2818 \cos(2 \theta) + 896 \cos(4 \theta) + 249 \cos(6 \theta) + 52 \cos(8 \theta) + 5 \cos(10 \theta)]/11520 .
\end{aligned}
\right.
\end{equation*}}
Figure~\ref{fig_3} shows a comparison between the numerical solution of Eq.~\eqref{rho_theta} and the fourth-order asymptotic expansion of $\rho$ for the representative values of $n = \{3, 5, 10, \infty\}$. As evident from the figure, the asymptotic expansion captures remarkably well the evolution of $\rho$ with $\theta$, and will therefore be used in the present study whenever an explicit formula for the dependence of $\rho$ on $\theta$ will be expedient.

\begin{figure}[!th]
\centering
\includegraphics[width=1.0\textwidth]{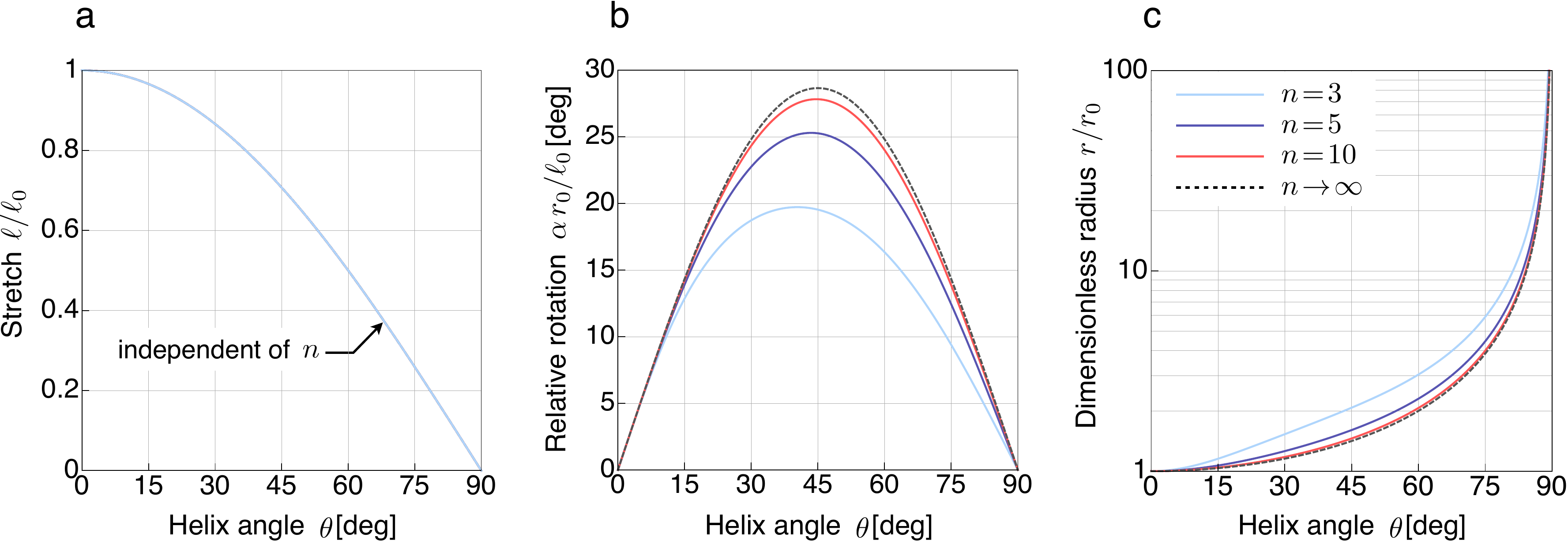}
\caption{(a) Stretch $\ell/\ell_0$, (b) relative rotation $(r_0/\ell_0)\alpha$ between the end sections of the assembly, and (c) dimensionless radius $r/r_0$ as a function of helix angle $\theta$ for representatives values of $n\!=\!\{3,5,10,\infty\}$. As for the graph in (a), notice that the stretch is independent of the number of strips. In fact, the vertical stretch simply reads $\ell/\ell_0 = \cos\theta$,  see Eq.~\eqref{height}. As regards the relative rotation $(r_0/\ell_0)\alpha$ shown in (b), notice that this is non-monotonic in $\theta$, see also the deformed configurations reported in the next figure.}
\label{fig_4}
\end{figure}

We next examine some features of the kinematics of helical assemblies, which will be useful in the next section. As the helix angle $\theta$ increases, the structural system shortens. In fact, from the parameterization of the rod's axis of Eq.~\eqref{rod_axis} we obtain that the current height of the assembly is given by
\begin{equation}
\ell = \ell_0 \cos\theta .
\label{height}
\end{equation}
In general, shortening of the assembly is accompanied by the relative rotation between its end sections of amount
\begin{equation}
\alpha =  \frac{\ell_0}{r} \sin\theta ,
\label{rotation}
\end{equation}
a relation that can be easily inferred from the argument of the trigonometric functions in Eq.~\eqref{rod_axis}. We report such kinematic features in Fig.~\ref{fig_4} in dimensionless form. In particular, Fig.~\ref{fig_4}a shows the vertical stretch $\ell/\ell_0$ as a function of helix angle $\theta$, whereas Fig.~\ref{fig_4}b reports on the relative rotation $\alpha\, r_0/\ell_0$ for the representative values of $n\!=\! \{3,5,10,\infty\}$. Notice that, while the stretch ratio is independent of $n$ and monotonic in $\theta$, this is not the case for the relative rotation. For completeness, Fig.~\ref{fig_4}c reports the evolution of the dimensionless radius $r/r_0$ with helix angle $\theta$.
%
%

\begin{figure}[!th]
\centering
\includegraphics[width=0.95\textwidth]{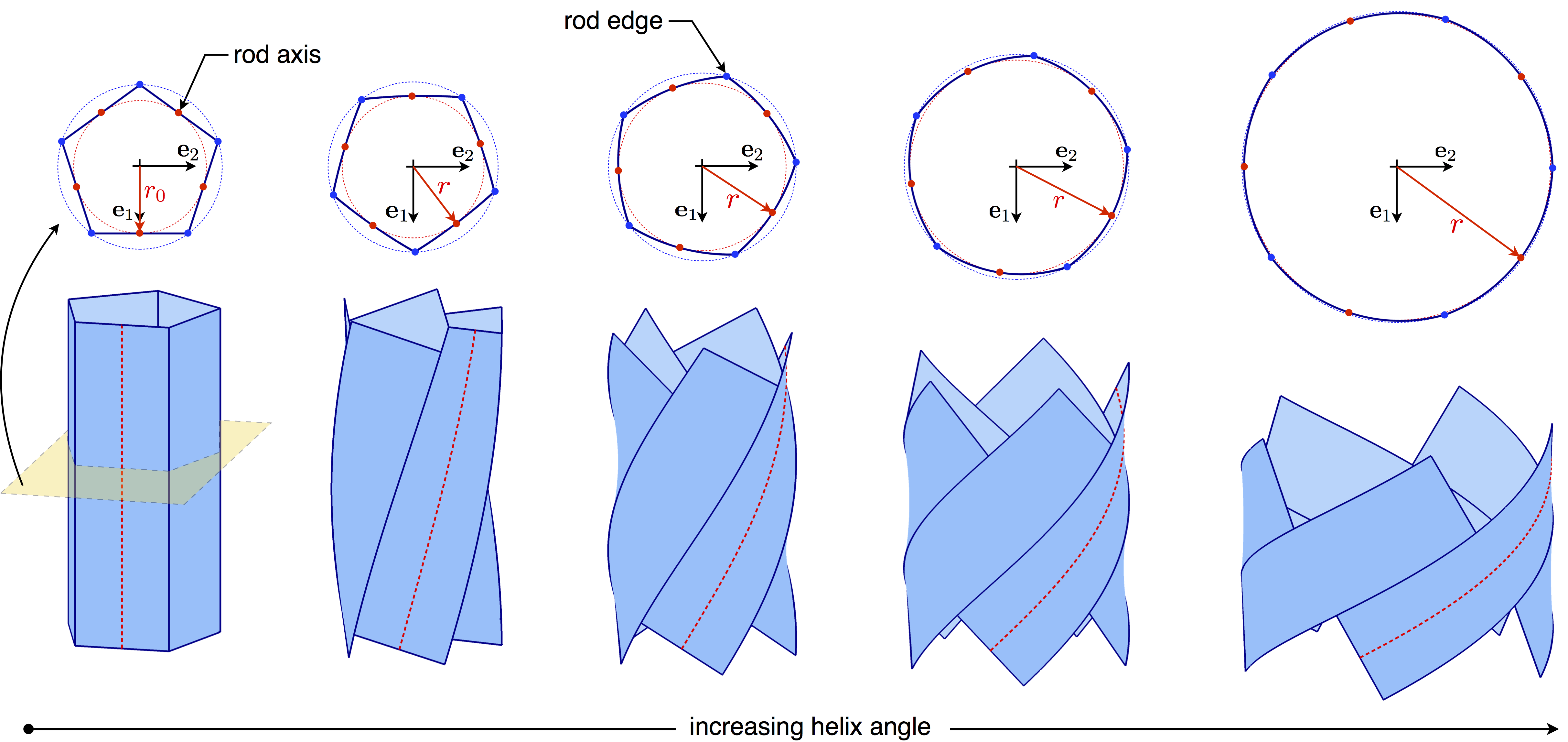}
\caption{Deformed shapes of a structural assembly of five rods along with respective, horizontal cross section computed at $s=\ell_0 / 2$. Specifically, deformed shapes are reported for increasing values of the helix angle $\theta\!=\!\{0,15,30,45,60\}$\,deg. Upon deformation, initially straight rods bend and twist into cylindrical helices, such that the height of the assembly decreases as $\theta$ increases. The decrease in height is accompanied by significant lateral expansion of the assembly, and by the relative rotation between its extremities. As for the cross sections, notice the peculiar geometry of the system, with rods axes belonging to the inscribed (red dashed) circle and rods edges belonging to the circumscribed (blue dashed) circle. Straight segments connect the rods edges in the undeformed configuration, such that the relevant cross section is a regular polygon. As the helix angle increases, those segments remain tangent to the inscribed (red dashed) circle but acquire significant curvature.}
\label{fig_5}
\end{figure}

To further highlight the main features of the assembly kinematics, Fig.~\ref{fig_5} shows deformed shapes of a structural system with $n=5$ along with respective, horizontal cross section computed for $s=\ell_0 / 2$.
In the figure, deformed shapes are reported for increasing values of the helix angle $\theta\!=\!\{0,15,30,45,60\}$\,deg. As regards the cross sections, notice the peculiar geometry of the system, with rods axes belonging to the inscribed (red dashed) circle and rods edges belonging to the circumscribed (blue dashed) circle. Straight segments connect the rods edges in the undeformed configuration, such that the relevant cross section is a regular polygon. By increasing the helix angle, those segments remain tangent to the inscribed (red dashed) circle but acquire significant curvature as a result of the kinematic assumptions behind the nonlinear rod theory used here.
%
%

With the aim of exploring the mechanical response of helical assemblies under external loading, we now introduce strain measures which are appropriate for the rod theory here employed. The three directors $\bm{d}_k(s)$, $k\!=\!\{1,2,3\}$, form an orthonormal basis for $\mathbb{E}^3$. Hence, we can define a vector-valued function $\bm{u}(s)$ such that
\begin{equation}
\bm{d}_k' = \bm{u} \times \bm{d}_k ,
\label{strains}
\end{equation}
where a prime denotes differentiation with respect to the arc-length parameter $s$ \citep{antman_2005}. Notice that, for simplicity, we have dropped the superscript \lq$i$' for the rod's number. We next represent $\bm{u}(s)$ in the basis $\bm{d}_k(s)$, {\it i.e.}~we write $\bm{u}(s) = u_k(s) \bm{d}_k(s)$, where we assume summation over the repeated index $k$.
The scalar-valued functions $u_k(s)$ are strain measures corresponding to the rod motion of Eq.~\eqref{rod_axis}. In particular, $u_1(s)$ and $u_2(s)$ measure flexure, whereas $u_3(s)$ measures twist.
By exploiting Eq.~\eqref{strains} we obtain
\begin{equation}
\bm{d}_1' = u_3 \bm{d}_2 - u_2 \bm{d}_3 , ~~~~~~~ \bm{d}_2' = - u_3 \bm{d}_1 + u_1 \bm{d}_3 , ~~~~~~~ \bm{d}_3' = u_2 \bm{d}_1 - u_1 \bm{d}_2 ,
\label{dirder}
\end{equation}
%
such that the three strain components can be readily computed as
\begin{equation}
u_1 = \bm{d}_2' \cdot \bm{d}_3 = - \bm{d}_3' \cdot \bm{d}_2 , ~~~~~~~ u_2 = \bm{d}_3' \cdot \bm{d}_1 = - \bm{d}_1' \cdot \bm{d}_3 , ~~~~~~~ u_3 = \bm{d}_1' \cdot \bm{d}_2 = - \bm{d}_2' \cdot \bm{d}_1 .
\end{equation}
%

Recall that, by construction, $\bm{d}_1(s) = \bm{n}(s)$, $\bm{d}_2(s) = \bm{b}(s)$, and that $\bm{d}_3(s) = \bm{t}(s)$. Consequently, we conclude that for the helically arranged rods of Fig.~\ref{fig_2}b-right
\begin{equation}
u_1 = 0 , ~~~~~~~ u_2 = \kappa = \frac{\sin^2\theta}{r} , ~~~~~~~ u_3 = \tau  = \frac{\sin\theta \cos\theta}{r} ,
\label{strains_helix}
\end{equation}
%
such that the strain measures correspond to the curvature $\kappa$ and torsion $\tau$ of the circular helices $\bm{x}^i(s)$, functions of $\theta$ only, since so is the radius $r$. 

To fully characterise the kinematics of the helical assembly, we derive now the explicit expression for the relative sliding $\delta$ between two adjacent rods. With reference to Fig.~\ref{fig_2}c, this can be readily computed as the length of any rod edge corresponding to an increase of $\delta s$ in the arc-length parameter. By differentiating $\bm{x}^i_{\pm}(s)$ with respect to $s$, we obtain the unit  vectors tangent to the rod's edges, that is
\begin{equation}
\bm{t}^i_\pm(s) = \frac{2}{\sqrt{4 + h^2 \tau^2}} \left[\bm{t}^i(s) \mp (h/2) \tau \bm{n}^i(s)\right] ,
\end{equation}
where use was made of Eqs.~\eqref{dirder} and \eqref{strains_helix}. Hence, for the computation of the relative sliding $\delta$ we write
\begin{equation}
\delta = \! \int_s^{s+\delta s} \lvert \bm{t}^i_\pm(\zeta) \rvert \mbox{d}\zeta = \frac{\delta s}{2} \sqrt{4 + h^2 \tau^2} ,
\label{sliding}
\end{equation}
such that, in general, $\delta$ will differ from $\delta s$. In fact, the quantity $\delta/\delta s$ is a measure of the \lq geometric stretch' undergone by the rod edges as a consequence of helix torsion $\tau$. This geometric effect highlights an inherent limitation of the rod model employed in the present study. However, such stretching effect will be negligible for $n$ sufficiently large, when $h$ is limited for a given, referential radius $r_0$.

\subsection{Elasticity}

We will now focus on the mechanical characterization of the rods composing helical structures such as that shown in Fig.~\ref{fig_5}. Let us denote by $\bm{m}(s)$ the resultant torque acting on a rod at the cross section singled out by $s$. We next introduce the scalar functions $m_k(s) = \bm{m}(s) \cdot \bm{d}_k(s)$, $k\!=\!\{1,2,3\}$. Of course, $m_1$ and $m_2$ are the bending moments about $\bm{d}_1$ and $\bm{d}_2$, respectively, whereas $m_3$ is the twisting moment about $\bm{d}_3$. In the present study, we will focus on assemblies of elastic rods, such that $m_k = \widehat{m}_k(u_k, u_k^*)$, where $u_k^*$ is a spontaneous strain. In particular, we will assume linear constitutive relations such that
\begin{equation}
m_1 = B_1(u_1 - u_1^*), ~~~~~~~ m_2 = B_2(u_2 - u_2^*), ~~~~~~~  m_3 = T(u_3 - u_3^*) ,
\end{equation}
where $B_1$ and $B_2$ are bending stiffnesses, whereas $T$ is the torsional stiffness of the cross section.

%
\begin{figure}[!th]
\centering
\includegraphics[width=1.0\textwidth]{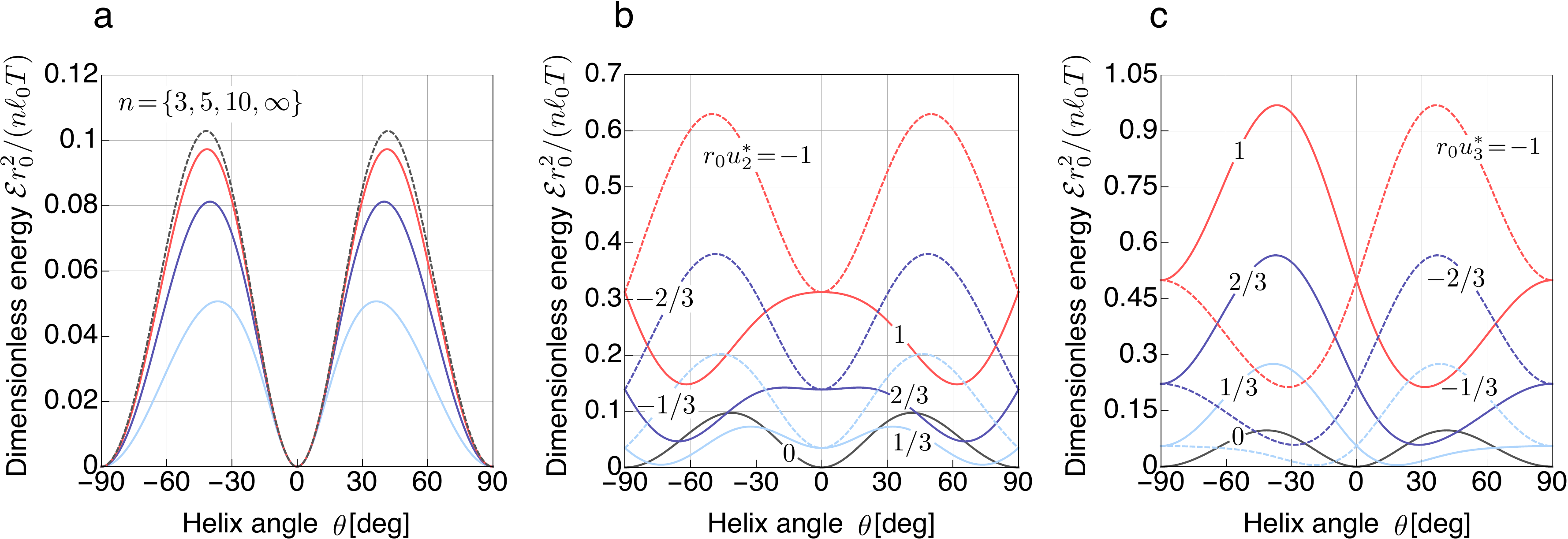}
\caption{(a) Dimensionless strain energy $\mathcal{E}r_0^2/(n\ell_0 T)$ in the absence of spontaneous strains as a function of helix angle $\theta$ and for $\beta_2 = 0.625$. Results are reported for $n\!=\!3$ (light blue), $n\!=\!5$ (blue), $n\!=\!10$ (red), and $n\!=\!\infty$ (black dashed). (b) Effect of spontaneous bending $u_2^*$ on the energy landscape for the representative case of $n\!=\!10$, and for $r_0u_2^*\!=\! \pm\{1/3,2/3,1\}$. (c) Effect of the spontaneous twisting $u_3^*$ on the energy landscape for the representative case of $n\!=\!10$, and for $r_0u_3^*\!=\! \pm\{1/3,2/3,1\}$.}
\label{fig_6}
\end{figure}
%

Our goal is to characterize the mechanical response under external loading of structural assemblies such as that shown in Fig.~\ref{fig_5}. To this aim, we next derive the elastic strain energy of the mechanical system. By accounting for the number of rods $n$, the strain energy reads
\begin{equation}
\mathcal{E} = \frac{n}{2} \int_0^{\ell_0} \left[B_1 (u_1 - u_1^*)^2 + B_2(u_2 - u_2^*)^2 + T(u_3  - u_3^*)^2 \right] \mbox{d}s ,
\end{equation}
which, in view of Eqs.~\eqref{strains_helix} and taking $u_1^* = 0$, becomes
\begin{equation}
\mathcal{E} = \ell_0 \frac{n}{2} \left[B_2(\kappa - u_2^*)^2 + T(\tau - u_3^*)^2 \right] ,
\label{energy_1}
\end{equation}
or in dimensionless form (per unit length and unit rod)
\begin{equation}
\mathcal{E} \frac{r_0^2}{n \ell_0 T} = \frac{1}{2} \left[\beta_2(r_0 \kappa - r_0 u_2^*)^2 + (r_0 \tau - r_0 u_3^*)^2 \right] ,
\label{energyadm_1}
\end{equation}
%
where the parameter $\beta_2 = B_2/T$. Equation~\eqref{energyadm_1} clearly shows that the spontaneous strains $u_k^*$ are in fact spontaneous curvatures, such that $r_0 u_k^*$ are dimensionless quantities.

We report in Fig.~\ref{fig_6}a the energy landscape in dimensionless form for the simple case in which spontaneous strains are absent, that is for $u_1^* = u_2^* = u_3^* = 0$. As previously done, results are shown as a function of $\theta$ and for the representative values of $n\!=\!\{3,5,10,\infty\}$ and with $\beta_2 = 0.625$.
Instead, the effect on the energy landscape of the spontaneous bending $r_0u_2^*\!=\! \pm\{1/3,2/3,1\}$ and of the spontaneous twisting $r_0u_2^*\!=\! \pm\{1/3,2/3,1\}$ is examined in Fig.~\ref{fig_6}b and in Fig.~\ref{fig_6}c, respectively, for the case of $n=10$ and for $\beta_2 = 0.625$\footnote{
For the simple case of a rectangular cross section of height $h$ and base $t$, the bending and torsional stiffnesses read
\begin{equation*}
B_2 = E \frac{t^3 h}{12}, ~~~~~~ T = G \frac{t^3 h}{3} ,
\end{equation*}
where $E$ and $G$ are the Young's and the shear modulii, respectively. Hence, for an elastic isotropic material
\begin{equation*}
\beta_2 = \frac{B_2}{T} = \frac{E}{4 G} = \frac{1 + \nu}{2} ,
\end{equation*}
such that the ratio of bending stiffness to torsional stiffness depends only on the Poisson's ratio $\nu$. For the representative values of $\nu\!=\!\{0, 0.25, 0.5\}$ one obtains $\beta_2\!=\!\{0.5, 0.625, 0.75\}$.
}. These figures portray a highly nonlinear energy landscape, which can be strongly tuned by the spontaneous curvature and twist of individual rods. Since this landscape determines the mechanics of the assembly, we anticipate a complex mechanical behavior for this system, even in the restricted case of cylindrical deformations.

\section{Mechanical response of helical assemblies under external loading and internal actuation}

We turn now our attention to the computation of the mechanical response under external loading of helical assemblies whose kinematics and energetics was introduced in the previous section. We do so by neglecting boundary layer effects at the extremities of the structural system and inter-strip friction. We explore separately the simple case of pure axial loading along the cylinder axis, the case of pure torque about that axis, and the case of a pressure acting upon the lateral surface of the assembly.

Given our working assumptions, {\it i.e.}~absence of inter-strip friction and boundary layer effects, computation of the mechanical response for the loading conditions discussed above can be achieved by exploiting the power balance for the structural system. In our context, this simply dictates the power of the external forces to balance the rate of change of the elastic strain energy. 

\subsection{Case of pure axial loading along the cylinder axis}

We start by exploring the response of the system for the case of pure axial loading along the cylinder axis $\mb{e}_3$, see Fig.~\ref{fig_2}b. We first notice that both the elastic strain energy, $\mathcal{E}(\theta)$, and the current length of the assembly, $\ell(\theta)$, are explicit functions of the helix angle $\theta$, recall Eqs.~\eqref{height} and \eqref{energy_1}. Hence, we compute the magnitude of the axial force $N$ corresponding to any configuration of the assembly, as parameterized by the helix angle, by writing
\begin{equation}
N = \frac{\mbox{d}\mathcal{E}}{\mbox{d}\ell} = \frac{\mbox{d}\mathcal{E}}{\mbox{d}\theta} \left(\frac{\mbox{d}\ell}{\mbox{d}\theta}\right)^{\!\! -1} \!.
\end{equation}
%

Substitution of \eqref{energy_1} in the equation above and elementary computations, not reported here for brevity, lead to
%
\begin{equation}
\begin{aligned}
N\frac{r_0^2}{n \, T} =
& \, \frac{1}{\rho \sin\theta} \left[\beta_2 r_0 u_2^* \sin(2\theta) + r_0 u_3^* \cos(2\theta)\right] - \frac{1}{\rho^2} \frac{\mbox{d} \rho}{\mbox{d} \theta} \left[\beta_2 r_0 u_2^* \sin\theta + r_0 u_3^* \cos \theta\right] + \\[3mm]
+ & \, \frac{\cos\theta}{\rho^2} \left[\left(\beta_2 - 1\right)\cos(2\theta) - \beta_2\right] + \frac{\sin\theta}{\rho^3} \frac{\mbox{d} \rho}{\mbox{d} \theta} \left[\left(\beta_2 - 1\right)\sin^2 \theta + 1\right] ,
\end{aligned}
\label{resp_N}
\end{equation}
an equation that provides, in dimensionless form and per unit rod, the axial force acting upon the system for a given configuration. We recall that the dimensionless radius $\rho$ is function of $\theta$ and of the rods number $n$, whereas $\beta_2 = B_2/T$ and $u_2^*$ and $u_3^*$ are spontaneous strains.

\begin{figure}[!th]
\centering
\includegraphics[width=1.0\textwidth]{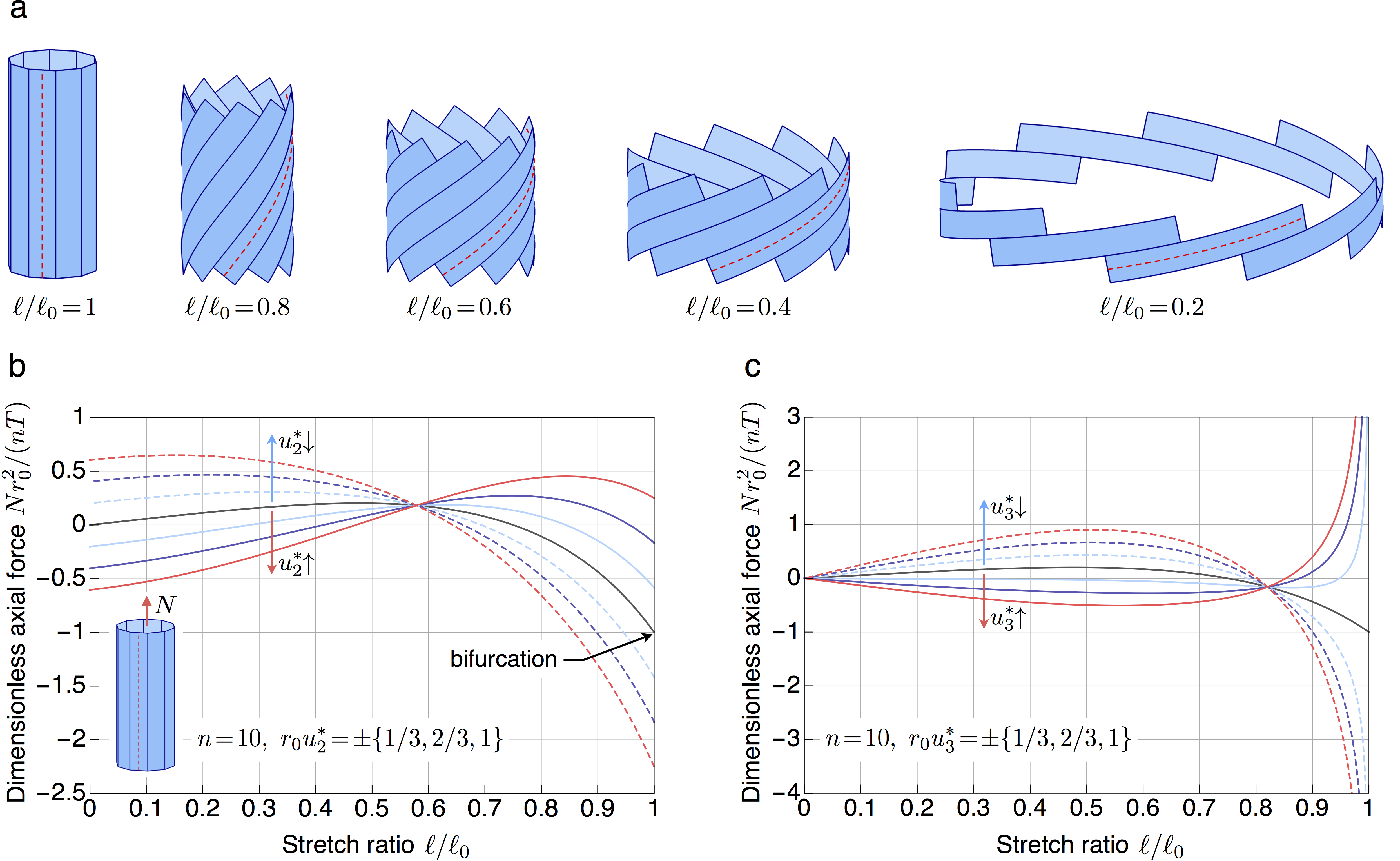}
\caption{(a) Deformed configurations of a structural assembly with $n\!=\!10$ rods at distinct levels of axial stretch, namely $\ell/\ell_0\!=\!\{1,0.8,0.6,0.4,0.2\}$. Notice the lateral expansion concomitant with the axial shortening. Effect of spontaneous bending (b) and of spontaneous twisting (c) on the axial-force versus vertical-stretch mechanical response for the representative case of $n=10$ and $\beta_2=0.625$. Results are reported for $r_0 u_2^*\!=\!\pm\{1/3,2/3,1\}$ and for $r_0 u_3^*\!=\!\pm\{1/3,2/3,1\}$ in (b) and (c), respectively, with (dashed) solid curves corresponding to (negative) positive values of the spontaneous curvatures. The black solid curve corresponds to the reference conditions of $u_2^* = u_3^* = 0$.}
\label{fig_7}
\end{figure}

We report in Fig.~\ref{fig_7}a deformed configurations for an assembly comprising $n=10$ rods at distinct values of the vertical stretch, namely $\ell/\ell_0\!=\!\{1,0.8,0.6,0.4,0.2\}$. Notice the lateral expansion of the system concomitant with the vertical shortening. Furthermore, we explore in Fig.~\ref{fig_7}b and in Fig.~\ref{fig_7}c the effect upon the equilibrium path of spontaneous bending and of spontaneous twisting of the rods, respectively. Specifically, results are shown for $r_0 u_2^*\!=\!\pm\{1/3,2/3,1\}$ and for $r_0 u_3^*\!=\!\pm\{1/3,2/3,1\}$ in Fig.~\ref{fig_7}b and in Fig.~\ref{fig_7}c, respectively, and compared with the case in which spontaneous curvatures are absent, black solid curve computed for $u_2^* = u_3^* = 0$. Interestingly, the equilibrium paths are non-monotonic functions of the vertical stretch, such that the external, axial force ranges from negative (compressive) to positive (tensile) values depending on the configuration of the system and on the spontaneous curvatures.

As for the results of Fig.~\ref{fig_7}b, the equilibrium curves always depart from finite values of the axial force $N$ evaluated at $\ell/\ell_0 = 1$, either positive or negative depending upon the sign and magnitude of $u_2^*$. This behaviour suggests the occurrence of a bifurcation of the equilibrium path at $\ell/\ell_0=1$, that is when the system is in the reference configuration and the rods are straight and vertically aligned. We further explore this peculiar behaviour by considering the total potential energy $\mathcal{P}$ of the structural system, which, taking into account the loading condition, reads
\begin{equation}
\mathcal{P} = \mathcal{E} - N\ell_0(\cos\theta - 1) ,
\end{equation}
such that its second-order Taylor expansion in the neighborhood of $\theta = 0$, that is for $\ell/\ell_0 = 1$, yields to
\begin{equation}
\mathcal{P} = \frac{n\, \ell_0 T}{2 r_0^2} \left[\beta_2 (r_0 u_2^*)^2 +  (r_0 u_3^*)^2 - 2 r_0 u_3^* \theta + \left(1 - 2\beta_2 r_0 u_2^* \right)\theta^2 \right] + \frac{N \ell_0}{2} \theta^2 + \mathcal{O}(\theta^3).
\label{toteng_2nd_axial}
\end{equation}
Next, equilibrium configurations of the system are sought for as extrema of the total potential energy. By neglecting higher order terms, differentiation of Eq.~\eqref{toteng_2nd_axial} with respect to $\theta$ leads to
\begin{equation}
N \frac{r_0^2}{n \, T} \theta = r_0 u_3^* - \left(1 - 2\beta_2 r_0 u_2^* \right)\theta .
\label{bal_2nd_axial}
\end{equation}

In the absence of spontaneous twist, {\it i.e.}~for $u_3^*=0$, it turns out that equilibrium of the structural assembly is possible either in the reference configuration of $\theta=0$ (or equivalently $\ell/\ell_0=1$) irrespective of the value of the axial force $N$, or in configurations with $\theta \neq 0$ (or equivalently $\ell/\ell_0 \neq 1$) when
\begin{equation}
N\frac{r_0^2}{n\,T} = -1 + 2\beta_2 r_0 u_2^* .
\end{equation}
As highlighted by the results of Fig.~\ref{fig_7}a, buckling of the system occurs at a load level which depends upon the spontaneous curvature of the rods $u_2^*$. Interestingly, the critical load is positive whenever $r_0 u_2^* > 1/\beta_2$, such that the system analysed in the present study is an example of a structure exhibiting tensile buckling \citep{zaccaria_2011, bigoni_2012}. As for the characterization of the reference configuration, this is stable for 
\begin{equation}
N\frac{r_0^2}{n\,T} > -1 +2\beta_2 r_0 u_2^* .
\end{equation}

We conclude our analysis by noticing that in the more general case of $u_3^* \ne 0$   
\begin{equation}
\lim_{\theta \to 0^{\pm}} N\frac{r_0^2}{n \, T} =  \pm \mbox{sign}(r_0 u_3^*) \infty ,
\end{equation}
a direct consequence of Eq.~\eqref{bal_2nd_axial}. Consistently with the equilibrium paths shown in Fig.~\ref{fig_7}c, whenever $u_3^* \ne 0$ an axial load of infinite magnitude is needed to keep the structural assembly in the reference configuration of $\ell/\ell_0 = 1$.

\subsection{Case of pure torque about the cylinder axis}

We continue our analysis by considering now the case of external loading by a torque about the cylinder axis $\mb{e}_3$, see Fig.~\ref{fig_2}b. As for the relative rotation between the extremities of the assembly, $\alpha(\theta)$, we recall that this is a function of the helix angle $\theta$ as given by Eq.~\eqref{rotation}.
\begin{figure}[!th]
\centering
\includegraphics[width=1.0\textwidth]{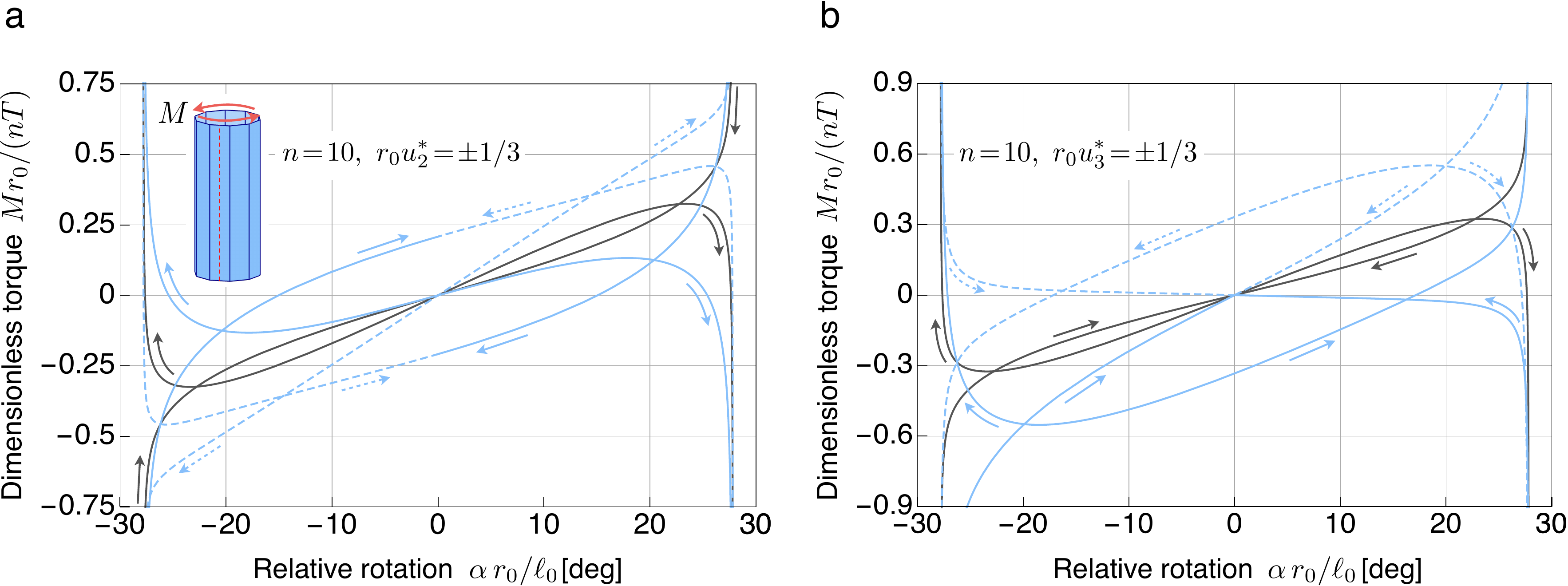}
\caption{Mechanical response under pure torsion of a structural assembly that comprises $n=10$ rods with $\beta_2 = 0.625$. Dimensionless torque $Mr_0/(n\,T)$ as a function of the relative rotation $\alpha\, r_0/\ell_0$ between the assembly extremities for the simple case of $u_2^* = u_3^* = 0$ (solid black curves) as compared with the two cases of (a) $r_0 u_2^* = \pm 1/3$ and of (b) $r_0 u_3^* = \pm 1/3$, respectively (solid and dashed light blue curves). Notice the arrows that highlight the loading path corresponding to increasing helix angle magnitude and the singular mechanical response of the system for $\alpha\,r_0/\ell_0 \simeq 28$\,deg.}
\label{fig_8}
\end{figure}
Therefore, computation of the magnitude of the external torque $M$ corresponding to any configuration of the system, as parameterized by the helix angle, is achieved by writing
\begin{equation}
M = \frac{\mbox{d}\mathcal{E}}{\mbox{d}\alpha}  =  \frac{\mbox{d}\mathcal{E}}{\mbox{d}\theta}\left(\frac{\mbox{d}\alpha}{\mbox{d}\theta}\right)^{\!\!-1} \!,
\label{powbal_M}
\end{equation}
%
that is, by equating the rate of change of the elastic strain energy with the power of the external torque. Similarly to the previous loading case, substitution of \eqref{energy_1} in the equation above yields
%
\begin{equation}
\begin{aligned}
M\frac{r_0}{n \, T} =
 & \, \frac{\sin\theta}{\rho\cos\theta - \displaystyle \frac{\mbox{d}\rho}{\mbox{d}\theta}\sin\theta} \left\{ \frac{\mbox{d}\rho}{\mbox{d}\theta} \left[\beta_2 r_0 u_2^* \sin\theta + r_0 u_3^* \cos\theta \right] - \frac{\rho}{\sin\theta}\left[\beta_2 r_0 u_2^* \sin(2\theta) + r_0 u_3^* \cos(2\theta) \right] \right.+ \\[2.5mm]
+ & \left. \frac{1}{2} \left[\left(\beta_2 + 1\right) \cos\theta - \left(\beta_2 - 1\right) \cos(3\theta)\right] - \frac{\sin\theta}{\rho} \frac{\mbox{d}\rho}{\mbox{d}\theta} \left[\left(\beta_2 - 1\right) \sin^2\theta + 1 \right] \right\} ,
\end{aligned}
\label{resp_M}
\end{equation}
an explicit expression for the dimensionless torque acting upon the structural assembly per unit rod.

Results from our analysis are reported in Fig.~\ref{fig_8} for the representative case of $n=10$ and $\beta_2 = 0.625$ in terms of the dimensionless torque $Mr_0/(n\,T)$ as a function of the relative rotation $\alpha\, r_0/\ell_0$ between the assembly extremities. In particular, Fig.~\ref{fig_8}a and Fig.~\ref{fig_8}b show the mechanical response of the system for the simple case of $u_2^* = u_3^* = 0$ (solid black curves) as compared with the two cases of $r_0 u_2^* = \pm 1/3$ and of $r_0 u_3^* = \pm 1/3$, respectively (solid and dashed light blue curves).

As previously noticed, the relative rotation between the end sections of the assembly is a non-monotonic function of the helix angle, recall Fig.~\ref{fig_4}b. Hence, irrespective of the spontaneous strains $u_2^*$ and $u_3^*$, the results of Fig.~\ref{fig_8} are characterized by two curves, corresponding to either increasing or decreasing values of $\alpha$. In order to distinguish between these two cases, curves in the figure are reported together with arrows (of appropriate colour and style) that select the loading path corresponding to increasing helix angle magnitude.
Notice that all the loading paths diverge for $\alpha\,r_0/\ell_0 \simeq 28$\,deg in the case of $n=10$ and $\beta_2=0.625$. This singular behaviour occurs at $\theta \simeq 45$\,deg and, from the mathematical standpoint, is due to the vanishing of the denominator in the factor outside the curly brackets of Eq.~\eqref{resp_M}, which also corresponds to the vanishing of the derivative of the relative rotation taken with respect to the helix angle, see Eq.~\eqref{rotation} and Fig.~\ref{fig_4}b. From a mechanical perspective, we interpret such singular response as follows. We mentioned that, for $\theta \simeq 45$\,deg, the derivative of the relative rotation $\alpha$ with respect to $\theta$ is null, whereas the derivative with respect to $\theta$ of elastic strain energy $\mathcal{E}$ is not, see Fig.~\ref{fig_6}. As a consequence, the power balance of Eq.~\eqref{powbal_M} requires the external torque to diverge in magnitude.

\subsection{Case of internal pressure acting upon the cylinder lateral surface}

We examined in the previous sections the two distinct cases of external loading by either a pure axial force or a pure torque about the cylinder axis. We conclude our study by exploring now the situation in which the structural assembly is internally actuated by a pressure acting upon its the lateral surface, a solution which is typically employed in a robotics context for the design of pneumatic actuators and artificial muscles. In particular, we assume a uniform pressure $p$ to act on a cylinder of radius $r$ and height $\ell$, such that its lateral surface area reads
\begin{equation}
S =  2 \pi r \, \ell = 2 \pi r \, \ell_0 \cos\theta .
\label{surface}
\end{equation}

As for the previous cases of external loading, we exploit the power balance to compute the pressure $p$ corresponding to any configuration of the assembly as parameterized by the helix angle $\theta$. More explicitly, we require the rate of change of strain energy to balance the power of the internal pressure and write
\begin{equation}
p = \frac{1}{S} \frac{\mbox{d}\mathcal{E}}{\mbox{d}r} =  \frac{1}{S} \frac{\mbox{d}\mathcal{E}}{\mbox{d}\theta} \left(\frac{\mbox{d}r}{\mbox{d}\theta}\right)^{\!\!-1} \!.
\label{powbal_p}
\end{equation}
%

By making use of Eq.~\eqref{energy_1}, the power balance of Eq.~\eqref{powbal_p} yields
%
\begin{equation}
\begin{aligned}
p\frac{r_0^4}{n \, T} =
& \, \frac{1}{2\pi \rho^2 \displaystyle \frac{\mbox{d}\rho}{\mbox{d}\theta}} \left\{ \frac{\sin\theta}{\rho}\frac{\mbox{d}\rho}{\mbox{d}\theta} \left[\beta_2 r_0 u_2^* \tan\theta + r_0 u_3^* \right] -  \left[2\beta_2 r_0 u_2^* \sin\theta + r_0 u_3^* \cos(2\theta)\sec\theta\right] \right. + \\[2.5mm]
+ & \, \left. \frac{\sin\theta}{\rho^2}\frac{\mbox{d}\rho}{\mbox{d}\theta} \left[\left(\beta_2 - 1\right)\sin\theta\cos\theta - \beta_2 \tan\theta\right] + \frac{\sin\theta}{\rho} \left[\beta_2 - \left(\beta_2 - 1\right)\cos(2\theta)\right] \right\},
\end{aligned}
\end{equation}
an equation that provides in dimensionless form and per unit rod the magnitude of the internal equilibrium pressure as a function of the helix angle $\theta$. 
\begin{figure}[!th]
\centering
\includegraphics[width=1.0\textwidth]{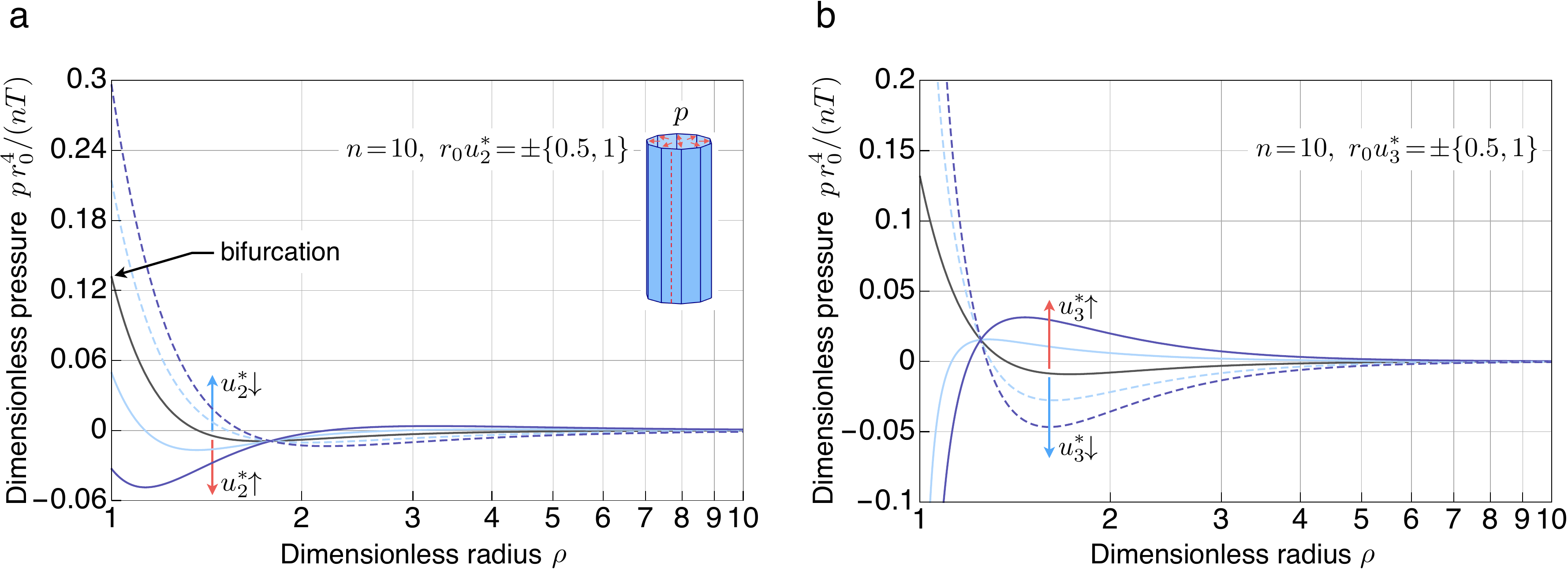}
\caption{Mechanical response under lateral pressure of a structural assembly comprising $n=10$ rods with $\beta_2 = 0.625$. Dimensionless pressure $p\,r_0^4/(n\,T)$ as a function of the dimensionless radius $\rho$ for the simple case of $u_2^* = u_3^* = 0$ (solid black curves) as compared with the two cases of (a) $r_0 u_2^*\!=\!\pm\{0.5,1\}$ and of (b) $r_0 u_3^*\!=\!\pm\{0.5,1\}$, respectively (solid and dashed curves).}
\label{fig_9}
\end{figure}

Results from our analysis are summarized in Fig.~\ref{fig_9} for the representative case of a structural assembly comprising $n=10$ rods with $\beta_2=0.625$. The figure reports on the mechanical response of the system in terms of the dimensionless pressure $p\,r_0^4/(n\,T)$ as a function of the dimensionless radius $\rho$. The effect upon the equilibrium path of spontaneous bending and of spontaneous twisting of the rods is explored separately in Fig.~\ref{fig_9}a and in Fig.~\ref{fig_9}b, respectively. Specifically, results are shown for $r_0 u_2^*\!=\!\pm\{0.5,1\}$ and for $r_0 u_3^*\!=\!\pm\{0.5,1\}$, and compared with the reference case in which spontaneous strains are absent (black solid curve).

Similarly to the case of pure axial loading, the mechanical system exhibits a bifurcation of the equilibrium path in its reference configuration, see Fig.~\ref{fig_9}a, such that the pressure versus radius equilibrium paths always depart from finite values of $p$ when $\rho = 1$. As for the total potential energy $\mathcal{P}$ of the system, this simply reads
\begin{equation}
\mathcal{P} = \mathcal{E} - p \, r_0 S (\rho - 1) ,
\end{equation}
while its second-order Taylor expansion in the neighbourhood of $\theta = 0$ yields\footnote{Notice that $\mbox{d}\rho(0)/\mbox{d}\theta = 0$ and that $\mbox{d}^2\!\rho(0)/\mbox{d}\theta^2 = 2 \sec^2(\pi/n) - 1$, as can be shown by first Taylor expanding both the LHS and the RHS of Eq.~\eqref{compatibility_2}, and then by balancing the terms of the same order in $\theta$.}
\begin{equation}
\mathcal{P} = \frac{n\,\ell_0 T}{2 r_0^2} \! \left[\beta_2 (r_0 u_2^*)^2 + (r_0 u_3^*)^2 - 2 r_0 u_3^*\theta + (1 - 2\beta_2 r_0u_2^*)\theta^2\right] - p\,\pi r_0^2 \ell_0 \! \left[2\sec^2\!\left(\frac{\pi}{n}\right) - 1\right]\! \theta^2 + \mathcal{O}(\theta^3) .
\label{toteng_2nd_pressure}
\end{equation}
Then, equilibrium configurations of the mechanical assembly in a neighborhood of $\theta=0$ are obtained by differentiation of Eq.~\eqref{toteng_2nd_pressure}, leading to
\begin{equation}
p \frac{r_0^4}{n\,T}\theta = \frac{r_0 u_3^* - (1 - 2\beta_2 r_0 u_2^*)\theta}{\displaystyle 2\pi\!\left[1 - 2\sec^2\!\left(\frac{\pi}{n}\right)\right]} \,.
\label{bal_2nd_pressure}
\end{equation}
It turns out that in the absence of spontaneous twist, that is for $u_3^* = 0$, equilibrium of the system is possible either in the reference configuration of $\theta = 0$ irrespective of the magnitude of pressure $p$, or in configurations with $\theta \ne 0$ when
\begin{equation}
p \frac{r_0^4}{n\,T} = \frac{- 1 + 2\beta_2 r_0 u_2^*}{\displaystyle 2\pi\!\left[1 - 2\sec^2\!\left(\frac{\pi}{n}\right)\right]} \, .
\end{equation}
As shown in Fig.~\ref{fig_9}a, buckling of the system occurs at a pressure level which critically depends upon the spontaneous curvature $u_2^*$. Intuitively, a positive pressure acting on the internal surface of the assembly induces a tensile circumferential stress which, in turns, leads to buckling by the relative sliding between adjacent rods, a mechanism that resembles that investigated by \cite{zaccaria_2011}. As regards the reference configuration of $\theta = 0$, this is stable for 
\begin{equation}
p \frac{r_0^4}{n\,T} < \frac{- 1 + 2\beta_2 r_0 u_2^*}{\displaystyle 2\pi\!\left[1 - 2\sec^2\!\left(\frac{\pi}{n}\right)\right]} \, .
\end{equation}

As for the correct interpretation of the formula above, notice that the term $1-2\sec^2(\pi/n) < 0$ for $n \ge 3$. Consequently, in the more general case of $u_3^* \ne 0$ one finds that
\begin{equation}
\lim_{\theta \to 0^{\pm}} p\frac{r_0^4}{n \, T} =  \mp \mbox{sign}(r_0 u_3^*) \infty .
\end{equation}
Consistently with the equilibrium paths shown in Fig.~\ref{fig_9}b, whenever $u_3^* \ne 0$ a pressure of infinite magnitude is needed to keep the structural assembly in the reference configuration of $\rho = 1$, an observation that concludes our analysis.

\section{Conclusions and outlook}

Inspired by the architecture of the euglenoid pellicle, we have explored the kinematics and the mechanical response under external loading of helical assemblies of interlocking, elastic rods. By mimicking the biological template, deformation in such assemblies is achieved by the relative sliding between adjacent strips. For simplicity, we have restricted our study to the case of uniform inter-strip sliding, an assumption leading to cylindrical structural systems. In carrying out the analysis, we have neglected inter-strip friction and boundary layer effects at the two free ends of the cylinder, such that the equilibrium response of the structural systems could be obtained with simple analytical techniques based on energetic arguments. In particular, we have obtained explicit formulae for the prototypical cases of pure axial loading along the cylinder axis, pure torque about that axis, and pressure acting upon the lateral surface of the assembly. The simplifying assumptions we have made lead to the possibility of obtaining closed form solutions for the equilibrium configurations. These reveal a rich mechanical response that is easy to rationalize with the help of explicit formulas. 

While we are aware of the inherent limitations of our study, we believe that the results presented here highlight the remarkable potential in engineering and robotics applications of structural systems inspired by the euglenoid pellicle. Future studies by the authors will include the analysis of more general shapes and of the impact of inter-strip friction on the mechanical response. We also plan to validate our theoretical findings by carrying out physical experiments on biomimetic structures realized by means of multi-material additive manufacturing techniques.

\section*{Acknowledgments}
ADS and GN acknowledge the support of the European Research Council (AdG-340685-MicroMotility). MA acknowledges the support of the European Research Council (CoG-681434), the Generalitat de Catalunya (2017-SGR-1278 and ICREA Academia prize for excellence in research). The authors would like to congratulate Prof. Norman Fleck on the occasion of his 60th birthday.

\section*{References}

\bibliography{bibliography}

\end{document}